\documentclass[11pt,a4paper]{article}
\pdfoutput=1
\usepackage{jheppub}
\usepackage{mathrsfs,graphicx,rotating,amsmath,amsfonts,mathtools,booktabs,wasysym,caption}
\usepackage{hyperref}
\usepackage{slashed}
\usepackage[table,xcdraw,dvipsnames]{xcolor}
\usepackage{graphicx}
\usepackage{bbold}
\usepackage[utf8x]{inputenc}
\usepackage[english]{babel}
\usepackage{multirow,multicol}
\usepackage{epstopdf}
\usepackage{bbm}
\usepackage{changepage}
\usepackage{appendix}
\usepackage{systeme}
\usepackage{libertine}
\usepackage{braket}
\usepackage{wasysym}
\usepackage{empheq}
\usepackage{cancel}
\usepackage{enumitem}
\usepackage{mathrsfs}
\usepackage{lmodern}
\usepackage{tabularx}
\usepackage{multicol}
\usepackage{color}
\usepackage{mathtools}
\usepackage{verbatim}
\usepackage{amssymb}
\usepackage{amsfonts}

\newcommand{\be}{\begin{equation}}
\newcommand{\ee}{\end{equation}}
\newcommand{\bea}{\begin{eqnarray}}
\newcommand{\eea}{\end{eqnarray}}

\newcommand{\mc}{\mathcal}

\newcommand{\beqa}{\begin{eqnarray}}
\newcommand{\eeqa}{\end{eqnarray}}

\setcitestyle{square}

\title{Kination, Meet Kasner: On The Asymptotic Cosmology of String Compactifications}
\author[a]{Fien Apers,} \author[a]{Joseph P. Conlon,}\author[a]{Martin Mosny}  \author[b]{and Filippo Revello}
\affiliation[a]{Rudolf Peierls Centre for Theoretical Physics\\ Beecroft Building, Clarendon Laboratory, Parks Road, University of Oxford, OX1 3PU, UK}
\affiliation[b]{Institute for Theoretical Physics\\ Utrecht University,
Princetonplein 5, 3584 CC Utrecht, The Netherlands}
\emailAdd{fien.apers@physics.ox.ac.uk}\emailAdd{joseph.conlon@physics.ox.ac.uk}\emailAdd{martin.mosny@physics.ox.ac.uk}\emailAdd{f.revello@uu.nl}
\abstract{We study runaway, kination-dominated epochs in string cosmology.
We show how the apparent classical decompactification runaway of the volume modulus, described by a kination epoch in the 4-dimensional EFT, 
can be uplifted to a classical Kasner solution in 10d in which the non-compact dimensions collapse towards a Big Crunch. This can also be generalised for arbitrary spacetime and compactification dimensions. We conclude with some comments on how this picture is modified by quantum effects, and the need for both dynamical and kinematical Swampland constraints.}
\begin{document}

\maketitle

\section{Introduction}

Cosmological solutions of string theory are important for several reasons. If a string cosmology ever described the actual early universe, its dynamical evolution is important for understanding the structure of this universe, our home. On a more general note, cosmological solutions are also interesting for the slightly more formal question of understanding the behaviours which are allowed, even in principle, within string theory.

One common feature of string theory vacua is the presence of runaway potentials towards an asymptotic limit \cite{Dine:1985he,BrusteinSteinhardt}. Such runaway potentials can arise from gaugino condensation in heterotic string theory, from non-perturbative superpotentials in type IIB compactifications, 
and also from perturbative corrections in type IIA, type IIB and other models.

The simplest asymptotic limits of moduli space are the large volume / weak coupling limits. In most
scenarios, the moduli that control the approach to these asymptotic limits are the K\"ahler (volume) and dilaton (string coupling) moduli.
The K\"ahler potential tends to be logarithmic in the asymptotic moduli, for example
\be
K = - 3 \ln \left( T + \bar{T} \right) - \ln \left( S + \bar{S} \right)
\ee
for a 1-modulus Calabi-Yau in the context of type IIB D3/D7 compactifications. The canonically normalised fields are also logarithmic in these moduli, $\Phi = \sqrt{\frac{3}{2}} \ln \tau$ and $\Psi = \sqrt{\frac{1}{2}} \ln s$, where $\tau = {\rm Re}(T)$ and $s = {\rm Re}(S) = \frac{1}{g_s}$. 
It follows that potentials with power-law expansions in $\tau^{-1}$ and $g_s$, as one would expect for physics originating in the 
perturbative $\alpha^{'}$ and $g_s$ expansions in string theory, are exponential when written in the canonical fields $\Phi$ and $\Psi$.

The development of precise scenarios of moduli stabilisation allows such dynamics to be studied in a robust setting rather than through \emph{ad hoc} assumptions. In our analysis, we first restrict to dynamics where only a single volume modulus controls the approach to the asymptotic limit, and regard the other fields (such as the dilaton) as rigid. This assumption is well motivated in the context of LVS \cite{Balasubramanian:2005zx}, where the moduli mass spectrum and generalised no-scale structure \cite{Conlon:2005ki} sees the volume modulus singled out as the unique light field, while the dilaton is flux-stabilised together with the other complex structure moduli at parametrically massive values \cite{Giddings:2001yu, Dasgupta:1999ss}. 

In this case, the simplest asymptotic potential (corresponding to a power-law expansion in $\tau^{-1}$, i.e. in inverse volume) is 
\be
V(\Phi) = V_0 \exp \left( - \lambda \Phi \right),
\ee
with $\lambda$ is a constant (although we can also consider double exponential potentials, $V(\Phi) \sim \exp \left( - \exp \left( \lambda \Phi \right) \right)$, arising from a purely non-perturbative superpotential $W = e^{-\alpha T}$). This familiar behaviour has also recently been reinterpreted in terms of the swampland de Sitter conjecture, at least in the asymptotic limit (see \cite{Obied:2018sgi}).

As long as $\lambda > \sqrt{2}$, it has long been understood that the scalar field evolution will result in kinetic energy domination and the scalar field rolling away to the asymptotic $\Phi \to \infty$ limit \cite{Wetterich:1987fm, Ferreira:1997hj, Copeland:1997et} (often referred to as a decompactification limit). Values of $\lambda < \sqrt{2}$ can give accelerated expansion but are hard to realise in string theory (see \cite{Garg:2018zdg,Hebecker:2019csg,ValeixoBento:2020ujr,Andriot:2020vlg, Cicoli:2021fsd,Cicoli:2021skd,Calderon-Infante:2022nxb, Andriot:2022brg} for recent analyses). As well as asymptotic limits, similar periods of kinetic energy domination can arise in string cosmology as transitory phases, as an intermediary epoch between the end of inflation and fields settling down into the present-day 
vacuum (see \cite{Conlon:2022pnx} for a recent study).

In the asymptotic limit, such kinetic-dominated runaways are normally viewed as phenomenologically undesirable, but not dangerous. The 
main point of this paper is to emphasise that this runaway behaviour conceals a more pathological aspect: its correct interpretation is as a 10-dimensional 
Kasner solution in the string frame, with the three external dimensions contracting and the six internal dimensions expanding. In an asymptotic limit, 
such a runaway solution is headed not towards decompactification: but instead towards a Big Crunch singularity in which the `non-compact' dimensions all collapse. 

The basic intuition for this is that, in a kinetic-dominated runaway, the potential becomes negligible for the dynamics. As the potential tends to be sourced by fluxes and branes within the compactified theory, in a limit of zero potential the dynamics of compactified effective field theories, at least in terms of the runaway field, should map onto sourceless cosmological solutions in the higher-dimensional theory. However, sourceless cosmological solutions do not allow for uniform expansion, but rather require the inhomogeneous behaviour of Kasner cosmologies.

The study of time-dependent cosmological backgrounds in string theory has a long history, including the study of the runaway epoch and generalisations of Kasner cosmologies (for example, see \cite{Antoniadis:1988vi,Mueller:1989in,
Arkani-Hamed:1999fet,Brandle:2000qp,Wands:2002ka,Giddings:2003zw,Giddings:2004vr} and for more recent work see \cite{Cicoli:2021fsd, Cicoli:2021skd, Rudelius:2022gbz}). However, this link of runaway stringy decompactification potentials to higher-dimensional Kasner cosmologies, leading to a subsequent Big Crunch in string frame, seems either not to have been made or to have so dropped out of general awareness that it needs restating.

The structure of this paper is as follows. In section \ref{sec4d} we demonstrate the relationship of Kasner and kination solution in 4d compactifications, first for one-modulus models and then in more general scenarios. In section \ref{secHigherDim} we consider the same question in compactifications to dimensions other than 4, finding that the same picture can be generalised for an arbitrary number of dimensions. In section \ref{secSwampland} we consider the relationship to the swampland and the question of which low-energy behaviours are dynamically possible within quantum gravity, before concluding.

\section{Kasner to Kination And Back Again}
\label{sec4d}

The Kasner solution \cite{Kasner:1921zz} is one of the oldest cosmological solutions of vacuum general relativity. In 10-dimensional spacetime, and splitting coordinates into $x_i$ (external) and $y_j$ (internal),
the Kasner metric becomes
\be
ds^2 = -dt^2 + \sum_{i=1}^3 t^{2p_i} dx_i^2 + \sum_{j=1}^{6} t^{2q_j} dy_j^2,
\ee
where
\be
\sum_i p_i + \sum_j q_j = 1, \qquad \sum_i p_i^2 + \sum_j q_j^2 = 1.
\ee
We now relate various forms of this solution to the dynamics of 4-dimensional effective field theories.

\subsection{EFTs with a single modulus}

In terms of 4-dimensional kination dynamics, we first consider the case where only a single volume modulus appears in the 4-dimensional EFT dynamics. We can then restrict the Kasner solution to homogeneous behaviour in the extra dimensions. In this case, two possible Kasner solutions exist.
The first has $p_i = -1/3$ and $q_j = 1/3$ (describing contraction in the 4-dimensional space and expansion in the compact space), while the latter has
$p_i = 5/9$ and $q_j = -1/9$ (with the 4-dimensional space expanding and the compact space contracting).

We first show that the former solution,
$$
ds^2 = -dt^2 + \sum_{i=1}^{3} t^{-2/3} dx_i^2 + \sum_{j=1}^{6} t^{2/3} dy_j^2,
$$
dimensionally reduces to the 4-dimensional runaway kination solution, with a scalar field rolling asymptotically towards infinity.

Treating the extra dimensions as compact, it follows from the Kasner metric that the extra-dimensional volume behaves as
\be
\left( \frac{{\rm Vol}(CY, t)}{{\rm Vol}(CY, t_0)} \right) \sim \left( \frac{t}{t_0} \right)^2.
\ee
As long as this volume is small, we must be able to view the physics from the perspective of 4-dimensional EFT, according to the standard procedures of 
dimensional reduction \cite{Polchinski:1998rr}. As applicable for LVS, we are regarding both the string coupling and other moduli as rigid and fixed with a high mass, allowing us to treat the only relevant rolling field as the volume (note that developments in moduli stabilisation are crucial for this, as otherwise it would be inconsistent to restrict the dynamics to only the volume mode and neglect all the other moduli). Dropping the non-dynamical and constant $g_s$ factors, the 4d Einstein frame metric relates to the 10d string frame metric as
\be
g_{4d, Einstein} = \mc{V} g_{4d, s},
\ee
where $\mc{V}$ is the internal volume in units of the string length $l_s$ and $g_{4d,s}$ is the 
restriction of the 10d string frame metric to the four large dimensions. We therefore have
\be
ds^2_{4d,E} = -t^2 dt^2 + t^{4/3} \left( dx_1^2 + dx^2 + dx_3^2 \right).
\ee
To put this in a more canonical form, we write $t_E = \frac{t^2}{2}$, to give
\be
ds^2_{4d,E} = - dt_E^2 + (2 t_E)^{2/3} \left( dx_1^2 +dx_2^2 + dx_3^2 \right),
\ee
with the growing scale factor $a(t_E) \sim t_E^{1/3}$ appropriate for a kinetic energy-dominated universe. 

In a 4-dimensional kinating phase, the evolution of the rolling scalar is
\be
\Phi(t_E) = \Phi(t_{E,0}) + \sqrt{\frac{2}{3}} M_P \ln \left( \frac{t_E}{t_{E,0}} \right).
\ee
Using the standard K\"ahler potential $K = -3 \ln \left( T + \bar{T} \right)$, it follows that the canonical field $\Phi = \sqrt{\frac{3}{2}} \ln {\rm Re}(T) = \sqrt{\frac{2}{3}} \ln \mc{V}$.
In the kination effective field theory, the compact volume therefore grows as $\mc{V} \sim t_E \sim t^2$ -- entirely consistent with the behaviour of the
10-dimensional Kasner solution.

We can also run this argument upwards, starting with the kination dynamics of a scalar field in 4-dimensions, with
\bea
\Phi & = & \Phi_0 + \sqrt{\frac{2}{3}} M_P \ln \left( \frac{t_E}{t_{E,0}} \right), \nonumber \\
ds^2 & = & -dt_E^2 + t_E^{2/3} \left( dx_1^2 + dx^2 + dx_3^2\right).
\eea
As above, our knowledge of the physical interpretation of moduli implies that the Calabi-Yau volume grows as $ {\rm Vol(CY)} \sim t_E/t_{E,0}$.
Together with our knowledge of the rules of dimensional reduction, this implies that the corresponding 10-dimensional string frame metric, restricted to 4d, behaves
as
\be
ds_{string}^2 = - \frac{d t_E^2}{t_E} + t_E^{-1/3} \left( dx_1^2 + dx_2^2 + dx_3^2 \right).
\ee
Performing a coordinate redefinition $t = t_E^{1/2}$, we obtain
\be
ds_{string}^2 = -dt^2 + t^{-2/3} \left( dx_1^2 + dx_2^2 + dx_3^2 \right),
\ee
which is the expected restriction of the Kasner solution to 4d.

This argument shows that pure runaway kination (i.e. a classical scalar field rolling to the infinite decompactification limit in an otherwise empty universe) should be interpreted as
representing the higher dimensional Kasner solution. The Kasner solution reveals a hidden problem with the asymptotics: the 3 ordinary dimensions are actually shrinking in string frame, and so the asymptotic limit is not 10-dimensional flat space but instead a singular crunch in the `non-compact' dimensions.

Note that this point was already lurking and implicitly present in the stringy interpretation of the 4d kination solution. Although this solution may appear 
unproblematic as the scale factor is growing as $a(t_E) \sim t_E^{1/3}$, this is within a frame where the 4-dimensional Planck scale, $M_P = 2.4 \times 10^{18} {\rm GeV}$, is fixed. As $ {\rm Vol(CY)} \sim t_E/t_{E,0}$, we see that the string scale actually behaves as $m_s \sim \frac{M_P}{\sqrt{\mc{V}}} \sim \frac{M_P}{\sqrt{t_E}}$ and so the string length behaves as $l_s \sim t_E^{1/2}$. So, although this kinating universe does have a growing scale factor, its fundamental scale $l_s$ is growing faster than the scale factor: given time, the fundamental scale much catch up with the expanding universe, with catastrophic consequences.

A similar analysis holds for the inwards kination solution, describing a roll in from large compact volumes. Here, the appropriate 10d Kasner metric is
$$
ds^2 = -dt^2 + \sum_{i=1}^{3} t^{10/9} dx_i^2 + \sum_{j=1}^{6} t^{-2/9} dy_j^2,
$$
and so  $\mc{V}_{CY} \sim t^{-2/3}$. The conversion to 4d Einstein frame gives
$$
ds_{4d,E}^2 = - t^{-2/3} dt^2 + t^{4/9} \left( dx_1^2 + dx_2^2 + dx_3^2 \right).
$$
Defining $t_E = t^{2/3}$, this metric becomes (up to numerical factors)
$$
ds_{4d,E}^2 = - dt_E^2 + t_E^{2/3} \left( dx_1^2 + dx_2^2 + dx_3^2 \right),
$$
which is the kination metric. In the 4-dimensional stringy effective field theory, this corresponds to cosmological evolution of the rolling volume field with
\be
\Phi(t) = \Phi(t_0) - \sqrt{\frac{2}{3}} M_P \ln \left( \frac{t_E}{t_{E,0}} \right).
\ee
It is easy to see that this does indeed give $\mc{V} \sim t_E^{-1} \sim t^{-2/3}$, as expected for consistency with the 10d Kasner solution.

\subsection{Multiple Moduli and Toroidal Compactifications}

It is instructive to extend this analysis to the case where multiple moduli control the extra-dimensional volume.
A useful example is when the Calabi-Yau volume can be split as $T^2 \times T^2 \times T^2$, with all tori dynamical. Cosmologically, 
this also describes fibred versions of LVS in which the fibre (torus) moduli are lighter than the volume modulus and so remain dynamical in the cosmological evolution \cite{Cicoli:2008gp}, even though the blow-up K\"ahler moduli are fixed along with the dilaton and complex structure moduli.

We therefore consider a toroidal configuration, with an internal manifold of $T^2 \times T^2 \times T^2$. In IIB compactifications, the 
chiral multiplets are defined in terms of 4-cycle volumes $\tau_i$, so we write
\be
\mc{V} = \sqrt{\tau_1 \tau_2 \tau_3},
\label{qeii}
\ee
(any numerical prefactor in Eq (\ref{qeii}) is irrelevant). The corresponding K\"ahler potential is
\be
K = - \ln \left( T_1 + \bar{T}_1 \right) - \ln \left( T_2 + \bar{T}_2 \right) - \ln \left( T_3 + \bar{T}_3 \right),
\ee
with $T_i = \tau_i + i c_i$, where $c_i$ is the RR axion (as these RR axions are not important for our subsequent dynamics, we shall not discuss them further). 
The kinetic terms are rendered canonical in terms of new fields $\phi_i = \sqrt{\frac{1}{2}} \ln \tau_i$. Within a kination-dominated runaway, the effective equations of motion for the $\phi_i$ fields are
\be
\ddot{\phi}_i + 3 H \dot{\phi}_i = 0,
\ee
and are solved by
\be
\phi_i(t) = \phi_{i}(t_{E,0}) + \alpha_i M_P \ln \left( \frac{t_E}{t_{E,0}} \right).
\ee
The Hubble scale is given by
\be
H = \frac{1}{M_P} \sqrt{\frac{1}{6} \left( \dot{\phi}_1^2 + \dot{\phi}_2^2 + \dot{\phi}_3^2 \right)} = \frac{1}{t_E} \sqrt{ \frac{\alpha_1^2 + \alpha_2^2 + \alpha_3^2}{6}}.
\ee
and so solutions of the equations of motion satisfy
\be
\alpha_1^2 + \alpha_2^2 + \alpha_3^2 = \frac{2}{3}.
\label{jpc}
\ee
The physical volume of each 4-cycle behaves as 
\be
\left( \frac{\tau_i}{\tau_{i,0}} \right) =  \left( \frac{t_E}{t_{E,0}} \right)^{\alpha_i \sqrt{2}},
\label{abc}
\ee
and so the overall volume behaves as
\be
\frac{\mc{V}}{\mc{V}_0} \sim \sqrt{\tau_1 \tau_2 \tau_3} = \left( \frac{t_E}{t_{E,0}} \right)^{\left( \alpha_1 + \alpha_2 + \alpha_3 \right)/\sqrt{2}}.
\label{volt}
\ee
To connect this to the Kasner solution, note that the volume of 4-cycle moduli $\tau_i$ relates to the volume of the 2-tori $t_i$ as
$\tau_i \sim \epsilon_{ijk} t_j t_k$, and so $\tau_1 \sim t_2 t_3$. It follows from Eq. (\ref{abc}) that the time-dependences of the 2-cycle volumes $t_i$ are
$$
t_1 \sim t_E^{\lambda_1}, \qquad t_2 \sim t_E^{\lambda_2}, \qquad t_3 \sim t_E^{\lambda_3},
$$
where
\bea
\lambda_1 & = & \frac{\alpha_2 + \alpha_3 - \alpha_1}{\sqrt{2}}, \\
\lambda_2 & = & \frac{\alpha_1 + \alpha_3 - \alpha_2}{\sqrt{2}}, \\
\lambda_3 & = & \frac{\alpha_1 + \alpha_2 - \alpha_3}{\sqrt{2}}.
\label{def}
\eea
As the exponents $\lambda_i$ directly tell us about the growth of the three 2-tori within the 6-dimensional space, they allow us to
re-interpret the 4-dimensional kination solution as a Kasner solution within the 10-dimensional theory. Using Eq. (\ref{volt}) to Eq. (\ref{def}) and rescaling
the 4d metric from 4d Einstein frame to 10d string frame, the required Kasner metric is
\bea
ds_{10d}^2 & = & \frac{-dt_E^2}{t_E^{\frac{\left( \alpha_1 + \alpha_2 + \alpha_3 \right)}{\sqrt{2}}}} + t_E^{\frac{2}{3} - \frac{\left( \alpha_1 + \alpha_2 + \alpha_3 \right)}{\sqrt{2}} }\left( dx_1^2 + dx_2^2 + dx_3^2 \right)  \\ 
& & + t_E^{\frac{\alpha_2 + \alpha_3 - \alpha_1}{\sqrt{2}}} \left( dy_1^2 + dy_2^2 \right) + t_E^{\frac{\alpha_1 + \alpha_3 - \alpha_2}{\sqrt{2}}} \left( dy_3^2 + dy_4^2 \right) + t_E^{\frac{\alpha_2 + \alpha_3 - \alpha_1}{\sqrt{2}}} \left( dy_5^2 + dy_6^2 \right). \nonumber
\eea
To turn this into canonical Kasner form, we write $t_E = t^{\frac{1}{1 - \frac{\left( \alpha_1 + \alpha_2 + \alpha_3 \right)}{2\sqrt{2}}}}$. The metric then becomes
\be
ds_{10d}^2 = - dt^2  + t^{2p} \left( dx_1^2 + dx_2^2 + dx_3^2 \right) + t^{2q_1} \left( dy_1^2 + dy_2^2 \right) + t^{2q_2} \left( dy_3^2 + dy_4^2 \right) +
t^{2q_3} \left( dy_5^2 + dy_6^2 \right),
\ee
where the Kasner exponents are given by
\bea
p & =  & \frac{2 \sqrt{2} - 3 \left( \alpha_1 + \alpha_2 + \alpha_3 \right)}{6\sqrt{2} - 3 \left( \alpha_1 + \alpha_2  + \alpha_3 \right) }, \\
q_1 & = & \frac{ \left( \alpha_2 + \alpha_3 - \alpha_1 \right) }{2 \sqrt{2} - \left( \alpha_1 + \alpha_2 + \alpha_3 \right) }, \\
q_2 & = & \frac{ \left( \alpha_1 + \alpha_3 - \alpha_2 \right) }{2 \sqrt{2} - \left( \alpha_1 + \alpha_2 + \alpha_3 \right) }, \\
q_3 & = & \frac{ \left( \alpha_1 + \alpha_2 - \alpha_3 \right) }{2 \sqrt{2} - \left( \alpha_1 + \alpha_2 + \alpha_3 \right) }.
\eea
It can be verified that these exponents indeed satisfy the Kasner conditions,
\bea
3p + 2 \left( q_1 + q_2 + q_3 \right) & = & 1, \\
3p^2 + 2 \left( q_1^2 + q_2^2 + q_3^2 \right) & = & 1,
\eea
provided that the condition of Eq. (\ref{jpc}) is satisfied, namely $ \alpha_1^2 + \alpha_2^2 + \alpha_3^2 = \frac{2}{3}$ -- thus showing that the 4-dimensional kination solution can indeed be interpreted as a higher-dimensional Kasner cosmology in string frame.

It is again straightforward to reverse this argument, so that it now runs from the 10-dimensional Kasner solution into a 4-dimensional kination solution. 
We start with a 10-dimensional Kasner solution,
\be
ds_{10d}^2 = - dt^2  + t^{2p} \left( dx_1^2 + dx_2^2 + dx_3^2 \right) + t^{2q_1} \left( dy_1^2 + dy_2^2 \right) + t_s^{2q_2} \left( dy_3^2 + dy_4^2 \right) +
t^{2q_3} \left( dy_5^2 + dy_6^2 \right).
\ee
As the volume of the internal dimensions is $\mc{V}_{CY} \sim t_s^{2\left( q_1 + q_2 + q_3 \right) }$, the 4d Einstein frame metric is
\be
ds_{4d,E}^2 = -t_s^{2\left( q_1 + q_2 + q_3 \right)} dt_s^2  + t_s^{2p + 2\left( q_1 + q_2 + q_3 \right)} \left( dx_1^2 + dx_2^2 + dx_3^2 \right).
\ee
Using the coordinate redefinition $t_E = t_s^{1 + \left( q_1 + q_2 + q_3 \right) }$, we obtain
\be
ds_{4d, E}^2 = - dt_E^2 + t_E^{\frac{2p + 2\left( q_1 + q_2 + q_3 \right)}{1 + \left( q_1 + q_2 + q_3 \right) } } \left( dx_1^2 + dx_2^2 + dx_3^2 \right),
\ee
which indeed reduces to the kination metric
\be
ds_{4d, E}^2 = - dt_E^2 + t_E^{2/3} \left( dx_1^2 + dx_2^2 + dx_3^2 \right)
\ee
provided the exponents satisfy the Kasner condition $3p + 2 \left( q_1 + q_2 + q_3 \right) = 1$.

\subsection{Perturbations of Kasner}

If a 4-dimensional runway kination solution is equivalent to a higher-dimensional Kasner solution, then it should also be the case that perturbations of this Kasner solution should have an interpretation as perturbations within the 4-d kination solution. In the case of the kination solution, it is long understood that a radiation perturbation of the kination solution will grow relative to the background, ending up in a tracker solution  \cite{Wetterich:1987fm,Copeland:1997et,Ferreira:1997hj}.
A growing perturbation of the Kasner solution, which places energy in a mode which corresponds, under dimensional reduction, to radiation degrees of freedom in the 4-dimensional theory, should then map under dimensional reduction to the growth of radiation in a kination background.

While we leave a full analysis including wave-like behaviour for future work, we discuss here some simple perturbations of the Kasner solution.
Hence, we consider the perturbed Kasner metric 
\begin{equation}\label{perturbed2}
\begin{split}
ds^2 = ds_{\text{kasner}}^2 & + \frac{2\epsilon t^{2/3}}{\sqrt 6}\sum_{\substack{i=1,2,3\\j={4, \dots, 9}}} h_{ij}(t)dx_idy_{j-3} \\
& + 2\epsilon^2 t^{2/3} \sum_{\substack{i,j=1,2,3\\i < j}} h_{ij}^2(t)dx_idx_j + \epsilon^2 t^{2/3} \sum_{i=1,2,3} h_{ii}^2(t)dx_i^2.
\end{split}
\end{equation}
This involves turning on components of the metric (the mixed $g_{xy}$ components) which correspond on toroidal dimensional reduction to vector fields, 
$A_{\mu}$.\footnote{Although we note that on a full Calabi-Yau the absence of 1-cycles implies that no $U(1)$ gauge field would arise in this manner.}
We can solve the vacuum Einstein equations to first order in $\epsilon$ if
\begin{equation}
 h_{ij}'(t) = 0,
\end{equation}
for all $i$ and $j$, whereas if we further restrict to
\begin{equation}
h_{ij}(t)=1
\end{equation}
the solution is then exact to all orders.

Note that locally this perturbation is equivalent to the Kasner metric, as they are related through the coordinate transformation
\begin{equation}
x_i \rightarrow x_i, \ \ \ \ \ \ \ y_j \rightarrow y_j+\tfrac{\epsilon}{\sqrt 6}(x_1+x_2+x_3).
\end{equation} 
However, such a coordinate transformation violates the global structure of the metric. In particular, it mixes the 
compact and noncompact coordinates and destroys the coordinate periodicity of the compactification structure
\begin{equation}
x_i \sim x_i + 2\pi R_i, \ \ \ \  \ \ \ y_j \sim y_j + 2\pi R_j.
\end{equation}
In contrast, the perturbed metric is defined within the compactification coordinates which satisfy the periodicity conditions, allowing for a 
simple and physical interpretation of compactification on reduction to the 4-dimensional theory.

Reducing to 4d as previously, we now get a string frame metric
\begin{equation}
ds_{\text{string}}^2 = -dt^2 + t^{-2/3}(dx_1^2 +dx_2^2+dx_3^2)+\epsilon^2 t^{2/3}(dx_1 +dx_2+dx_3)^2.
\end{equation}
 In Einstein frame, we have
\begin{equation}\label{rad}
ds_{4,E}^2 = t^2 d s^2_{\text{string}} = -d\tau^2 + (2\tau)^{2/3}(dx_1^2 + dx_2^2+dx_3^2) + \epsilon^2 (2\tau)^{4/3} (dx_1 + dx_2+dx_3)^2,
\end{equation}
with $\tau = t^2/2$.
To first order in $\epsilon$ the scale factor is $a(\tau) = \tau^{1/3}$ while the energy density associated to \eqref{rad} is 
\begin{equation}
\rho = \tfrac{1}{24\pi}\tau^{-2} + \tfrac{2^{5/3}}{24\pi}\epsilon^2 \tau^{-4/3} \sim a^{-6} + \epsilon^2 a^{-4},
\end{equation}
which is the energy density applicable to kination with a small but growing radiation fraction, consistent with the expected behaviour in 4d EFT. 
Although we leave a full analysis of perturbations to future work, this provides evidence that perturbed Kasner solutions indeed capture the growth of radiation in a kination background, viewed from a 4d perspective.

\section{Kasner and Kination in Higher Dimensions}
\label{secHigherDim}
We next investigate whether similar behaviours hold for compactifications from $D$ to $(d+1)$ spacetime dimensions, where $D$ and $d$ can take any values (provided that $D > d+1>1$).
\subsection{Kasner to Kination}
In a $(d+1)$-dimensional FLRW spacetime with vanishing potential, the Friedmann equation is (See Appendix \ref{app:B} for details)
\be
\label{eq:f}
\left( \frac{\dot{a}}{a} \right)^2 = H^2 = \frac{2}{d(d-1)M_{P,d+1}^2} \rho = \frac{\dot{\phi}^2}{M_{P,d+1}^2 d(d-1)},
\ee
where $M_{P,d+1}$ is the $(d+1)$-dimensional Planck mass.
The scalar equation of motion is
\be
\ddot{\phi}+ d H \dot{\phi} =0.
\ee
Together with Eq. \eqref{eq:f}, this implies
\begin{equation}\label{eq:ked}
\phi(t_E) = \phi_0(t_E) +  \sqrt{\frac{d-1}{d}} M_{P,d+1} \log \left( \frac{t_E}{t_{E,0}}\right), \quad \quad  \quad \rho(t_E) = \frac{d-1}{2d} \frac{M_{P,d+1}^2}{ t_E^2},
\end{equation}
with the scale factor and Hubble scale behaving as
\bea
H(t_E) & = & \frac{1}{d\, t_E }, \\
 a(t_E) & = & \left( \frac{t_E}{t_{E,0}} \right)^{1/d}, 
 \eea
 corresponding to a metric
 \be
ds_{E,d}^2 = dt_E^2 - t_E^{\frac{2}{d}}\sum_{i=1}^d dx_i^2.
\label{kinationddim}
\ee
We ask whether Eq. (\ref{kinationddim}) can be obtained through compactification of a $D$-dimensional Kasner solution. For a generic $D-$dimensional spacetime, a Kasner solution compatible with an isotropic $d$-dimensional space has a metric 
\begin{equation}
\label{eq:k}
ds^2 = dt^2- t^{2p} \sum_{i=1}^d dx_i^2-\sum_{j=1}^{D-d-1} t^{2q_i}dy_j^2,
\end{equation}
where the exponents must satisfy
\begin{equation}\label{eq:exp}
\begin{cases}
pd + \sum_{j=1}^{D-d-1} q_j =1,\\
p^2 d + \sum_{j=1}^{D-d-1} q_j^2 =1.\\
\end{cases}
\end{equation}
For homogeneous internal dimensions, $q_i = q$, with the two solutions
\begin{equation}
\bar{p}_{\pm}= \frac{1 \pm \sqrt{ \frac{(D - 2) (D - d - 1)}{d}} }{D-1}  \quad \quad \bar{q}_{\pm}= \frac{1\mp\sqrt{\frac{(D-2)d}{D-d-1}}}{D-1}.
\label{dfg}
\end{equation}
To see whether \eqref{eq:k} can reproduce $(d+1)$-dimensional kination, as in section \ref{sec4d} we move to Einstein frame via 
$g_{\mu \nu, E} = \left( \mathcal{V}_{D-d-1} \right)^{2/(d-1)}\, g_{\mu \nu,s}$ (See Appendix \ref{app:A}). As the higher dimensional volume scales as
\begin{equation}
\mathcal{V}_{D-d-1} \sim t^{\sum_{j=1}^{D-d-1} q_j},
\end{equation}
the $(d+1)$-dimensional part of the metric becomes
\be
ds_{4d,E}^2  =   t^{\frac{2(1-pd)}{d-1}} dt^2- t^{\frac{2(1-p)}{d-1}} \sum_{i=1}^d dx_i^2,
\ee
where we have used \eqref{eq:exp}. With the substitution
\begin{equation}
t = \left( \frac{d-pd}{d-1} t_E \right)^{\frac{d-1}{d-pd}}, 
\end{equation}
we obtain
\begin{equation}
ds_{4d,E}^2 = dt_E^2-t_E^{\frac{2}{d}}\sum_{i=1}^d dx_i^2,
\end{equation}
matching with \eqref{kinationddim}. 

Furthermore, as happens in $d=3$, the kinating solution is unstable to perturbations (such as the presence of initial radiation), which drive it towards a tracker solution in any dimension $d>1$. Indeed, radiation and kination energy densities redshift as\footnote{Matter, which redshifts as $a(t_E)^{-d}$, can also drive kination towards a tracker solution, at a rate faster than radiation. In practice though, it is hard to have states behaving as matter when the Hubble scale is very high, for example after inflation.}
\begin{equation}\label{eq:cu}
\rho_{{\text{rad}}} (t_E)= \frac{\rho_{{\text{rad}},0}}{a(t_E)^{d+1}} \quad \quad {\text{and}} \quad \quad \rho_{{\text{kin}}} (t_E) = \frac{\rho_{{\text{kin}},0}}{a(t_E)^{2d}}
\end{equation}
respectively, so that that the latter will eventually catch up with the former. More precisely, one can repeat the analysis of \cite{Copeland:1997et}, and find that kination along a sufficiently steep exponential potential, in the presence of radiation, always evolves to an attractor solution where the energy densities have fixed ratios.  The details are presented in Appendix \ref{app:B}.

\subsection{Kination to Kasner}
We remark that, so far, we haven't imposed the requirement that the scalar field giving rise to kination be identified with the extra dimensional volume(s). 
This aspect is less developed in $d \neq 3$ spatial dimensions, as the nature of light scalar fields depends on the moduli stabilisation scenario, and these have been most studied for compactifications to 3 spatial dimensions.
Since the volume has a fixed time dependence during kination, this leads to an additional condition that must be satisfied in order for the uplift to work.\footnote{It would also be interesting to consider the cases where fields other than the volume, e.g. the dilaton, are driving kination.}

For simplicity, let us firstly treat the case where a single scalar is controlling the size of the extra dimensions, and all the exponents $q_j=q$ are 
equal (analogous to the LVS case considered in section \ref{sec4d}). In the string frame, the volume scales as
\begin{equation}\label{eq:vs}
\mathcal{V}_S \sim t^{ \sum_{j=1}^{D-d+1}q_j } = t^{1-pd} \sim t_E^{\frac{(1-pd)(d-1)}{d-pd}} 
\end{equation}

If there is a single modulus controlling the size of the volume, the canonically normalised scalar is (See Appendix \ref{app:A})
\begin{equation}\label{eq:vt}
\Phi =  \sqrt{\frac{D-2}{(D-d-1)(d-1)}}  M_{P,d+1}\log \mathcal{V}_S, \quad  \mathcal{V}_S \sim t_E^{(d-1) \sqrt{\frac{D-d-1}{d(D-2)}}}.
\end{equation} 
If $\Phi$ is the field responsible for kination, this should agree with \eqref{eq:vs}. Indeed, the two exponents are equal if
\begin{equation}
p^2d(D-1)-2 p d+2+d-D =0,
\end{equation}
which is implied by \eqref{eq:exp}. Therefore, a kinating volume modulus can be
uplifted to a Kasner solution for any total spacetime dimension $D$ and compactification dimension $d+1$.

\section{Applications to the Swampland}
\label{secSwampland}

The physics we have described involves evolution that, from a 4-dimensional perspective, appears as an ordinary kinating scalar field. However, the trajectory of the field leads towards a Big Crunch and so cannot be continued indefinitely within the full theory. This is reminiscent of swampland restrictions on 
allowed behaviours within 4d effective field theories within quantum gravity UV completions.

\subsection{What goes wrong in the effective field theory?}

In a kination phase in 4d EFT, the field moves through roughly one Planckian distance every Hubble time
\be
\Delta \Phi = \sqrt{ \frac{2}{3}} M_P \ln \left( \frac{t_E}{t_{E,0}} \right).
\ee
We have seen in section \ref{sec4d} that, even classically, this field profile cannot be sustained indefinitely due to the presence of a Crunch singularity.
But what goes wrong? How can we see this from a 4-dimensional perspective and how does the 4-dimensional effective field theorist know about his or her imminent demise? When measuring scales, the effective field theorist sees $\frac{H}{m_{KK}} \to 0 $ and $\frac{H}{m_{\Psi}} \to 0$ along the kination roll, an apparent indicator that they are being rapidly transported to the Isles of the Blessed in which all couplings are weak and all volumes are large. But, in fact, they are doing Turkey Effective Field Theory: and Christmas (or Thanksgiving) is coming in the form of the Crunch.

In a theory such as LVS, the light moduli are the volume and the axion; all other moduli obtain large masses and so can consistently be integrated out of the low-energy effective field theory. Within the cosmological kination solution, the masses of these other moduli satisfy $m_{\Psi} > H$ and so will not be subject to gravitational particle production. Although it is true that a tower of particles becomes progressively lighter as the kinating field evolves to the boundary of moduli space, consistent with the Swampland Distance Conjecture \cite{Klaewer:2016kiy}, it is always the case that $m_{\text{tower}} > H$, and so the single-modulus description remains valid within low-energy effective field theory. The Transplanckian Censorship Conjecture (TCC) \cite{Bedroya:2019snp} is also satisfied,\footnote{As is a refined version proposed in \cite{Andriot:2022brg}.} as
\begin{equation}
\frac{a(t_E)}{a(t_{E,0})} \frac{H(t_E)}{M_P} = \left( \frac{t_E}{t_{E,0}} \right)^{1/3} \frac{1}{3 M_P t}  \ll 1
\end{equation}
for any times $t_E > t_{E,0}$.  Indeed, what is interesting is that the singularity arising en route to decompactification does not appear to trace back to a violation of the refined distance conjecture \cite{Ooguri:2006in,Klaewer:2016kiy,Baume:2016psm,Ooguri:2018wrx}, the EFT paradigm\footnote{We refer to Section 2.1 of \cite{Conlon:2022pnx} for a more detailed discussion on the validity of the EFT approach.} or any other swampland constraint. The behaviour involved instead appears harmless: a runway towards the weak-coupling, large-volume decompactification limit.

One 4-dimensional perspective on the underlying problem is that, although the scale factor $a(t_E) \sim t_E^{1/3}$ is growing in Einstein frame, the fundamental string length $l_s$ is growing faster ($l_s \sim t_E^{1/2}$) and the Crunch corresponds to the latter catching up with the former. However, to foresee this within 4-dimensional effective field theory, we need to know enough of the fundamental theory to know that a string length exists and what size it has.

One further 4-dimensional aspect to the question can be identified. Our analysis has focused on classical evolution and a classical runaway, starting with an empty universe with just a rolling modulus. In a quantum universe, even if initially empty, quantum particle production will occur and generate a small fraction of radiation (for example, for either the graviton or the massless axion always present in LVS). Along the kination roll, any initial amount of radiation, however small, will 
grow in relative importance (as $\rho_{\gamma} \sim a^{-4}$ while $\rho_{kination} \sim a^{-6}$) and the solution will evolve towards a tracker solution, 
moving away from the kination (and uplifted Kasner) dynamics. So, it is clear that quantum effects will disrupt this analysis and allow the kination phase only to last for a finite amount of time.

\subsection{Dynamics and Kinematics in the Swampland}

There is a large literature on swampland conjectures: bounds on the possible allowed forms of low-energy field theories consistent with quantum gravity. The vast majority of these, however, are \emph{kinematical} in nature: they are statements about the masses of towers of particles at different points in moduli space (for example, as in \cite{Grimm:2018ohb,Blumenhagen:2018nts,Grimm:2018cpv,Corvilain:2018lgw}) or the charge-to-mass ratio of particles at points in moduli space (as in the weak gravity conjecture \cite{Lee:2018urn,Lee:2018spm,Gendler:2020dfp,Bastian:2020egp}). 

Such conjectures do not address \emph{dynamical} evolution, except for a limited number of examples such as the TCC \cite{Bedroya:2019snp} or the Festina Lente \cite{Montero:2019ekk, Montero:2021otb} bound. The results of section \ref{sec4d} show that the `obvious' route to decompactification is anything but: rather than leading to decompactification into flat and empty 10-dimensional space, it instead produces a Big Crunch singularity within our three spatial dimensions. This motivates the question of whether there are restrictions on the nature of
dynamical trajectories towards the asymptotic regions of moduli space, such as the large-volume and weak-coupling limits? More in general, it would be interesting to come up with constraints on the allowed trajectories in field space, in a similar spirit to \cite{Calderon-Infante:2022nxb, Calderon-Infante:2020dhm}. Although our focus has been on evolution towards decompactified large-volume weak-coupling limits of moduli space, it would also be interesting to study an evolution towards other singular loci in moduli space, such as the ones analysed in \cite{Grimm:2019ixq}.

Another perspective on this is that in a kination phase, 
the fields moves through roughly a Planckian distance every Hubble time
\be
\Delta \Phi = \sqrt{ \frac{2}{3}} M_P \ln \left( \frac{t_E}{t_{E,0}} \right)
\ee
and so kination defines a field profile in spacetime in which the field vev alters by a Planckian distance every Hubble time. The essence of section \ref{sec4d} was that there are significant problems with such a profile. This suggests that fundamental constraints may exist on field profiles with parametrically
 trans-Planckian temporal field excursions. We leave, however, a precise formulation of such constraints for future work.

\section{Conclusions}

We have seen in this paper how the runaway, kination-dominated behaviour of scalar fields in string theory conceals an interesting higher-dimensional interpretation in terms of a 10-dimensional Kasner solution. This behaviour is interesting for both phenomenological and more formal reasons. From a phenomenological side, in various scenarios in string cosmology the universe goes through a kination phase between inflation and the final minimum (e.g. see  
\cite{Conlon:2022pnx} for a recent study). The higher-dimensional solution may give a novel perspective on the physics of this epoch.

From a more formal perspective, the kination runaway represents behaviour which appears satisfactory within 4-dimensional EFT, but conceals a pathology within the full theory. This is reminiscent of the swampland, and motivates the development of swampland constraints which are \emph{dynamical} in nature, corresponding to the time evolution of systems, rather than merely \emph{kinematic} statements about theories at particular points in moduli space.

Kination epochs in string cosmology are common, result in trans-Planckian field excursions, but are relatively understudied in the string cosmology literature. We regard the physics described in this paper as an example of why they are interesting and merit more detailed study.

\subsection*{Acknoweldgements}

We thank Prateek Agrawal, Ed Copeland, Thomas Grimm and Sirui Ning for discussions. FA is supported by the Clarendon Scholarship in partnership with the Scatcherd European Scholarship, Saven European Scholarship, and the Hertford College Peter Howard Scholarship. We acknowledge support from STFC grant ST/T000864/1.
The research of FR is partly supported by the Dutch Research Council
(NWO) via a Start-Up grant and a Vici grant. For the purpose of Open Access, the authors have applied a CC BY public copyright licence to any Author Accepted Manuscript version arising from this submission.

\appendix
\section{Dimensional reduction}\label{app:A}
We briefly review here how one can obtain the relation between the canonically normalised volume modulus in $d+1$ dimensions and the volume of the compact space. Although this can easily be achieved within a supergravity description (such as $4d$ $\mathcal{N}=1$ theories), direct dimensional reduction gives the result for any dimension. We start from a $D$-dimensional Einstein Hilbert action,
\begin{equation}
S_{EH} = \frac{1}{2 k_D^2} \int d^Dx  \, \sqrt{-\tilde{g}_{D}}  \mathcal{\tilde{R}}_{D},
\end{equation}
and decompose the higher dimensional metric as
\begin{equation}\label{eq:mh}
ds^2 = \tilde{g}_{\mu \nu} dx^{\mu} dx^{\nu}+ e^{2  \omega(x)} \tilde{g}_{m n} dy^{m} dy^{n} \quad \mu, \nu = 0,..,d \quad m,n= d+1,..,D-1.
\end{equation}
Therefore, the higher dimensional volume (in the string frame) is given by
\begin{equation}\label{eq:vn}
\mathcal{V}= e^{(D-d-1) \omega(x)}.
\end{equation}
Let us first perform a Weyl rescaling on the $D$-dimensional metric, through the field redefinition
\begin{equation}\label{eq:w1}
\tilde{g}_{ab}^{D} = e^{2 \omega(x)} g_{ab}^{D}.
\end{equation}
By use of standard identities in GR, the Lagrangian density can be rewritten as
\begin{equation}\label{eq:rsc}
\sqrt{-\tilde{g}_{D}}  \mathcal{\tilde{R}}_D  =  \sqrt{-g_D} e^{(D-2) \omega(x)} \left(\mathcal{R}_{D}-2 (D-1)  \nabla^2 \omega - (D-1)(D-2) \partial_{a} \omega \partial^{a} \omega \right),
\end{equation}
where all the contractions and covariant derivatives on the right-hand side are performed with respect to the new metric $g_{ab}^{D}$. Since the metric is now a product metric of two spaces, the higher dimensional Ricci scalar decomposes as the sum of the Ricci scalar in the two spaces, \emph{i.e.}
\begin{equation}
\mathcal{R}_{D} = \mathcal{R}_{d+1} + \mathcal{R}_{D-d-1},
\end{equation} 
with $g_{\mu\nu}^{d+1}$ the $(d+1)$-dimensional block of the full metric $g_{ab}^{D}$. In the bulk we are assuming the source-less Einstein equations to be satisfied, so $ \mathcal{R}_{(D-d-1)} = 0$.  We can then compactify to $d+1$ dimensions. Integrating out the KK modes from the spectrum, the action becomes
\bea
S_{EH} & = & \frac{M^2_{P,d+1}}{2} \int d^{d+1}x  \sqrt{-g_{d+1}} e^{(D-2) \omega(x)} \\
& &  \times \left(\mathcal{R}_{d+1} - 2 (D-1)  \nabla^2 \omega - (D-1)(D-2) \partial_{\mu} \omega \partial^{\mu} \omega \right). \nonumber
\eea
We have defined the $d$-dimensional (reduced) Planck mass as 
\begin{equation}
\frac{M^2_{P,d+1}}{2} = \frac{\tilde{\mathcal{V}}}{2 k_D^2},
\end{equation}
where ${\tilde{\mathcal{V}}}$ is the volume calculated with the metric $\tilde{g}_{mn}$. We can now perform a second Weyl rescaling of the $d+1$-dimensional metric only, 
\begin{equation}\label{eq:w2}
g_{\mu \nu}^{d+1} = e^{-\frac{2(D-2)}{d-1} \omega(x)} h_{\mu \nu},
\end{equation}
to make the Ricci scalar in $d+1$ dimensions canonically normalised. The latter will transform as dictated by \eqref{eq:rsc} under a scale transformation (with $D=d+1$), while the Laplacian transforms as
\begin{equation}
\nabla_g^2\, \omega = e^{-\frac{2(D-2)}{d-1} \omega(x)} \left(\nabla^2_h\, \omega -(D-2)  h^{\mu \nu} \partial_{\mu} \omega \partial_{\nu} \omega \right).
\end{equation}

Putting everything together, and turning any total derivatives into boundary terms, the action becomes
\begin{equation}
S_{EH} = \frac{M^2_{P,d+1}}{2} \int d^{d+1}x  \sqrt{-h} 
\left( \mathcal{R}_h- \frac{(D-2)(D-d-1)}{d-1}  \partial_{\mu} \omega \partial^{\mu} \omega  \right),
\end{equation}
so that the normalised volume scalar takes the form
\begin{equation}
\Phi = \sqrt{\frac{D-2}{(D-d-1)(d-1)}}   M_{P,d+1}\log \mathcal{V}.
\end{equation}
For $D=10$ and $d=3$, we recover $\Phi = \sqrt{\frac{2}{3}} M_{P}\log \mathcal{V}$. Combining Eqs. \eqref{eq:vn}, \eqref{eq:w1} and \eqref{eq:w2}, we also see that the Einstein frame metric $h_{\mu \nu}$ is related to the string frame metric $\tilde{g}_{\mu \nu}$ by
\begin{equation}
h_{\mu \nu} = \mathcal{V}^{\frac{2}{(d-1)}} \tilde{g}_{\mu \nu},
\end{equation}
which correctly reduces to $h_{\mu \nu} = \mathcal{V} \tilde{g}_{\mu \nu} $ for $d=3$.

\section{Tracker solution in arbitrary dimensions}\label{app:B}
We here show the existence of tracker solutions in arbitrary dimensions, for a kinating scalar field rolling along an exponential potential and in the presence of radiation. This confirms the expectations of Eq \eqref{eq:cu}. Our analysis will closely parallel that of \cite{Copeland:1997et}, generalising it to an arbitrary number of dimensions. We start from a $(d+1)$ dimensional FLRW metric,
\begin{equation}
ds^2 = -dt_E^2 +a(t_E)^2 \sum_{i=1}^d dx_i^2,
\end{equation}
and assume that (together with the kinating scalar) another cosmic fluid is present. For simplicity, we only consider the case of radiation, whose conservation law and equation of state are given by
\begin{equation}
\dot{\rho} + d H \left(\rho_{\gamma} + P_{\gamma} \right) = 0, \quad \quad \quad P_{\gamma} = \frac{\rho_{\gamma}}{d}.
\end{equation}
The $(d+1)$-dimensional Friedmann equations can be written as (see for example \cite{Nojiri:2002hz,Chen:2014fqa}) 
\begin{equation}
\dot{H}= - \frac{M^2_{P,d+1}}{d-1} \left(\rho_{\gamma}+P_{\gamma}+\frac{\dot{\Phi}^2}{2} \right),
\end{equation}
\begin{equation}
H^2 = \frac{2}{d(d-1)M_{P,d+1}^2} \left( \rho_{\gamma}+\frac{1}{2} \dot{\Phi}^2+V(\Phi) \right),
\end{equation}
and the equation of motion for the spatially homogeneous scalar is given by
\begin{equation}
\ddot{\Phi}+ d H \dot{\Phi} + \frac{\partial V}{\partial \Phi}=0.
\end{equation}
We define the variables
\begin{equation}
x = \frac{1}{ H M_{P,d}}\frac{\dot{\Phi}}{\sqrt{d(d-1) }}, \quad \quad y = \frac{1}{H M_{P,d}} \sqrt{\frac{2 V(\Phi) }{d(d-1)}},
\end{equation}
and take the specific choice of an exponential potential $V = V_0 e^{- \frac{\lambda \Phi}{M_{P,d+1}}}$. The Friedmann equations can then be recast as the dynamical system
\begin{equation}\label{eq:syk}
\begin{cases}
x' =- d x  + \lambda \sqrt{\frac{d (d - 1)}{4}} y^2 + 
 x  \left( d x^2 + \frac{d+1}{2} \left(1 - x^2 - y^2 \right) \right), \\
 y' =   -\lambda \sqrt{\frac{d (d - 1)}{4}} x y + y \left( d x^2 + \frac{d+1}{2} \left(1 - x^2 - y^2 \right) \right),  \\
 \end{cases}
\end{equation}
with the constraint that
\begin{equation}\label{eq:cst}
0 \leq \Omega_{\gamma} \leq 1, \quad \quad \Omega_{\gamma} \equiv \frac{2 \rho_{\gamma}}{d(d-1)H^2}  = 1-x^2-y^2 .
\end{equation}
The primed quantities in \eqref{eq:syk} represent derivatives with respect to $N = \log a$. Without loss of generality, we can assume $y \geq 0$, since sending $y\rightarrow - y$ is equivalent to inverting the sign of $t$. The system above always has the trivial critical points
\begin{equation}
(x,y)_1 = (0,0), \quad \quad \quad (x,y)_2= (1,0), \quad \quad \quad (x,y)_3= (-1,0),
\end{equation}
which correspond to the cases of first an empty universe and then kination (in either direction for $\Phi$) respectively. A linear stability analysis around those points shows they always have at least one runaway direction, and are thus an unstable node or a saddle point. There are also two non-trivial critical points for $d>1$, given by
\begin{equation}
(x,y)_4 = \left(\frac{\lambda}{2 }\sqrt{ \frac{d-1}{d}}, \frac{1}{2}\sqrt{4 -\lambda^2\frac{d-1}{d}}\right), \quad (x,y)_5 = \left(\frac{d+1}{\lambda \sqrt{d(d-1)}}, \frac{1}{\lambda}\sqrt{\frac{d+1}{d}} \right).
\end{equation} 
The first critical point is characterised by $\Omega_{\gamma,4}=0$, and is also present in the absence of radiation \cite{Rudelius:2022gbz}. It is a solution where the scalar field is not kinating ($y \neq 0$, so there is energy density in the potential), but there no radiation is present. It is an attractor for
\begin{equation}
\lambda^2 < \frac{2(d+1)}{(d-1)},
\end{equation}
and a saddle point otherwise (if it exists). For the second critical point, the constraint  \eqref{eq:cst} implies it only exists for
\begin{equation}
 \lambda^2 \geq \frac{2(d+1)}{(d-1)} \quad \quad {\text{as}}\quad \quad \Omega_{\gamma,5} = 1-\frac{2}{\lambda^2}\frac{d+1}{d-1}.
\end{equation}
Linearising \eqref{eq:syk} around $(x,y)_5$, one obtains that this last solution is a stable node for
\begin{equation}
\frac{2(d+1)}{(d-1)} < \lambda^2 < \frac{16(d+1)^2}{(9 + 7 d) (d - 1)}
\end{equation}
and a stable spiral for
\begin{equation}
\lambda^2 > \frac{16(d+1)^2}{(9 + 7 d) (d - 1)}.
\end{equation}
This analysis is in agreement with Table I of \cite{Copeland:1997et} for $d=3$.
\bibliography{biblist}

\providecommand{\href}[2]{#2}\begingroup\raggedright\begin{thebibliography}{10}

\bibitem{Dine:1985he}
M.~Dine and N.~Seiberg, \emph{{Is the Superstring Weakly Coupled?}},
  \href{https://doi.org/10.1016/0370-2693(85)90927-X}{\emph{Phys. Lett. B}
  {\bfseries 162} (1985) 299}.

\bibitem{BrusteinSteinhardt}
R.~Brustein and P.~J. Steinhardt, \emph{{Challenges for superstring
  cosmology}}, \href{https://doi.org/10.1016/0370-2693(93)90384-T}{\emph{Phys.
  Lett. B} {\bfseries 302} (1993) 196}
  [\href{https://arxiv.org/abs/hep-th/9212049}{{\ttfamily hep-th/9212049}}].

\bibitem{Balasubramanian:2005zx}
V.~Balasubramanian, P.~Berglund, J.~P. Conlon and F.~Quevedo,
  \emph{{Systematics of moduli stabilisation in Calabi-Yau flux
  compactifications}},
  \href{https://doi.org/10.1088/1126-6708/2005/03/007}{\emph{JHEP} {\bfseries
  03} (2005) 007} [\href{https://arxiv.org/abs/hep-th/0502058}{{\ttfamily
  hep-th/0502058}}].

\bibitem{Conlon:2005ki}
J.~P. Conlon, F.~Quevedo and K.~Suruliz, \emph{{Large-volume flux
  compactifications: Moduli spectrum and D3/D7 soft supersymmetry breaking}},
  \href{https://doi.org/10.1088/1126-6708/2005/08/007}{\emph{JHEP} {\bfseries
  08} (2005) 007} [\href{https://arxiv.org/abs/hep-th/0505076}{{\ttfamily
  hep-th/0505076}}].

\bibitem{Giddings:2001yu}
S.~B. Giddings, S.~Kachru and J.~Polchinski, \emph{{Hierarchies from fluxes in
  string compactifications}},
  \href{https://doi.org/10.1103/PhysRevD.66.106006}{\emph{Phys. Rev. D}
  {\bfseries 66} (2002) 106006}
  [\href{https://arxiv.org/abs/hep-th/0105097}{{\ttfamily hep-th/0105097}}].

\bibitem{Dasgupta:1999ss}
K.~Dasgupta, G.~Rajesh and S.~Sethi, \emph{{M theory, orientifolds and G -
  flux}}, \href{https://doi.org/10.1088/1126-6708/1999/08/023}{\emph{JHEP}
  {\bfseries 08} (1999) 023}
  [\href{https://arxiv.org/abs/hep-th/9908088}{{\ttfamily hep-th/9908088}}].

\bibitem{Obied:2018sgi}
G.~Obied, H.~Ooguri, L.~Spodyneiko and C.~Vafa, \emph{{De Sitter Space and the
  Swampland}},  \href{https://arxiv.org/abs/1806.08362}{{\ttfamily
  1806.08362}}.

\bibitem{Wetterich:1987fm}
C.~Wetterich, \emph{{Cosmology and the Fate of Dilatation Symmetry}},
  \href{https://doi.org/10.1016/0550-3213(88)90193-9}{\emph{Nucl. Phys. B}
  {\bfseries 302} (1988) 668}
  [\href{https://arxiv.org/abs/1711.03844}{{\ttfamily 1711.03844}}].

\bibitem{Ferreira:1997hj}
P.~G. Ferreira and M.~Joyce, \emph{{Cosmology with a primordial scaling
  field}}, \href{https://doi.org/10.1103/PhysRevD.58.023503}{\emph{Phys. Rev.
  D} {\bfseries 58} (1998) 023503}
  [\href{https://arxiv.org/abs/astro-ph/9711102}{{\ttfamily
  astro-ph/9711102}}].

\bibitem{Copeland:1997et}
E.~J. Copeland, A.~R. Liddle and D.~Wands, \emph{{Exponential potentials and
  cosmological scaling solutions}},
  \href{https://doi.org/10.1103/PhysRevD.57.4686}{\emph{Phys. Rev. D}
  {\bfseries 57} (1998) 4686}
  [\href{https://arxiv.org/abs/gr-qc/9711068}{{\ttfamily gr-qc/9711068}}].

\bibitem{Garg:2018zdg}
S.~K. Garg, C.~Krishnan and M.~Zaid~Zaz, \emph{{Bounds on Slow Roll at the
  Boundary of the Landscape}},
  \href{https://doi.org/10.1007/JHEP03(2019)029}{\emph{JHEP} {\bfseries 03}
  (2019) 029} [\href{https://arxiv.org/abs/1810.09406}{{\ttfamily
  1810.09406}}].

\bibitem{Hebecker:2019csg}
A.~Hebecker, T.~Skrzypek and M.~Wittner, \emph{{The $F$-term Problem and other
  Challenges of Stringy Quintessence}},
  \href{https://doi.org/10.1007/JHEP11(2019)134}{\emph{JHEP} {\bfseries 11}
  (2019) 134} [\href{https://arxiv.org/abs/1909.08625}{{\ttfamily
  1909.08625}}].

\bibitem{ValeixoBento:2020ujr}
B.~Valeixo~Bento, D.~Chakraborty, S.~L. Parameswaran and I.~Zavala, \emph{{Dark
  Energy in String Theory}},
  \href{https://doi.org/10.22323/1.376.0123}{\emph{PoS} {\bfseries CORFU2019}
  (2020) 123} [\href{https://arxiv.org/abs/2005.10168}{{\ttfamily
  2005.10168}}].

\bibitem{Andriot:2020vlg}
D.~Andriot, P.~Marconnet and T.~Wrase, \emph{{Intricacies of classical de
  Sitter string backgrounds}},
  \href{https://doi.org/10.1016/j.physletb.2020.136015}{\emph{Phys. Lett. B}
  {\bfseries 812} (2021) 136015}
  [\href{https://arxiv.org/abs/2006.01848}{{\ttfamily 2006.01848}}].

\bibitem{Cicoli:2021fsd}
M.~Cicoli, F.~Cunillera, A.~Padilla and F.~G. Pedro, \emph{{Quintessence and
  the Swampland: The Parametrically Controlled Regime of Moduli Space}},
  \href{https://doi.org/10.1002/prop.202200009}{\emph{Fortsch. Phys.}
  {\bfseries 70} (2022) 2200009}
  [\href{https://arxiv.org/abs/2112.10779}{{\ttfamily 2112.10779}}].

\bibitem{Cicoli:2021skd}
M.~Cicoli, F.~Cunillera, A.~Padilla and F.~G. Pedro, \emph{{Quintessence and
  the Swampland: The Numerically Controlled Regime of Moduli Space}},
  \href{https://doi.org/10.1002/prop.202200008}{\emph{Fortsch. Phys.}
  {\bfseries 70} (2022) 2200008}
  [\href{https://arxiv.org/abs/2112.10783}{{\ttfamily 2112.10783}}].

\bibitem{Calderon-Infante:2022nxb}
J.~Calder\'on-Infante, I.~Ruiz and I.~Valenzuela, \emph{{Asymptotic Accelerated
  Expansion in String Theory and the Swampland}},
  \href{https://arxiv.org/abs/2209.11821}{{\ttfamily 2209.11821}}.

\bibitem{Andriot:2022brg}
D.~Andriot, L.~Horer and G.~Tringas, \emph{{Negative scalar potentials and the
  swampland: an Anti-Trans-Planckian Censorship Conjecture}},
  \href{https://arxiv.org/abs/2212.04517}{{\ttfamily 2212.04517}}.

\bibitem{Conlon:2022pnx}
J.~P. Conlon and F.~Revello, \emph{{Catch-me-if-you-can: the overshoot problem
  and the weak/inflation hierarchy}},
  \href{https://doi.org/10.1007/JHEP11(2022)155}{\emph{JHEP} {\bfseries 11}
  (2022) 155} [\href{https://arxiv.org/abs/2207.00567}{{\ttfamily
  2207.00567}}].

\bibitem{Antoniadis:1988vi}
I.~Antoniadis, C.~Bachas, J.~R. Ellis and D.~V. Nanopoulos, \emph{{An Expanding
  Universe in String Theory}},
  \href{https://doi.org/10.1016/0550-3213(89)90095-3}{\emph{Nucl. Phys. B}
  {\bfseries 328} (1989) 117}.

\bibitem{Mueller:1989in}
M.~T. Mueller, \emph{{Rolling Radii and a Time Dependent Dilaton}},
  \href{https://doi.org/10.1016/0550-3213(90)90249-D}{\emph{Nucl. Phys. B}
  {\bfseries 337} (1990) 37}.

\bibitem{Arkani-Hamed:1999fet}
N.~Arkani-Hamed, S.~Dimopoulos, N.~Kaloper and J.~March-Russell, \emph{{Rapid
  asymmetric inflation and early cosmology in theories with submillimeter
  dimensions}},
  \href{https://doi.org/10.1016/S0550-3213(99)00667-7}{\emph{Nucl. Phys. B}
  {\bfseries 567} (2000) 189}
  [\href{https://arxiv.org/abs/hep-ph/9903224}{{\ttfamily hep-ph/9903224}}].

\bibitem{Brandle:2000qp}
M.~Brandle, A.~Lukas and B.~A. Ovrut, \emph{{Heterotic M theory cosmology in
  four-dimensions and five-dimensions}},
  \href{https://doi.org/10.1103/PhysRevD.63.026003}{\emph{Phys. Rev. D}
  {\bfseries 63} (2001) 026003}
  [\href{https://arxiv.org/abs/hep-th/0003256}{{\ttfamily hep-th/0003256}}].

\bibitem{Wands:2002ka}
D.~Wands, \emph{{String inspired cosmology}},
  \href{https://doi.org/10.1088/0264-9381/19/13/302}{\emph{Class. Quant. Grav.}
  {\bfseries 19} (2002) 3403}
  [\href{https://arxiv.org/abs/hep-th/0203107}{{\ttfamily hep-th/0203107}}].

\bibitem{Giddings:2003zw}
S.~B. Giddings, \emph{{The Fate of four-dimensions}},
  \href{https://doi.org/10.1103/PhysRevD.68.026006}{\emph{Phys. Rev. D}
  {\bfseries 68} (2003) 026006}
  [\href{https://arxiv.org/abs/hep-th/0303031}{{\ttfamily hep-th/0303031}}].

\bibitem{Giddings:2004vr}
S.~B. Giddings and R.~C. Myers, \emph{{Spontaneous decompactification}},
  \href{https://doi.org/10.1103/PhysRevD.70.046005}{\emph{Phys. Rev. D}
  {\bfseries 70} (2004) 046005}
  [\href{https://arxiv.org/abs/hep-th/0404220}{{\ttfamily hep-th/0404220}}].

\bibitem{Rudelius:2022gbz}
T.~Rudelius, \emph{{Asymptotic scalar field cosmology in string theory}},
  \href{https://doi.org/10.1007/JHEP10(2022)018}{\emph{JHEP} {\bfseries 10}
  (2022) 018} [\href{https://arxiv.org/abs/2208.08989}{{\ttfamily
  2208.08989}}].

\bibitem{Kasner:1921zz}
E.~Kasner, \emph{{Geometrical theorems on Einstein's cosmological equations}},
  \href{https://doi.org/10.2307/2370192}{\emph{Am. J. Math.} {\bfseries 43}
  (1921) 217}.

\bibitem{Polchinski:1998rr}
J.~Polchinski, \emph{{String theory. Vol. 2: Superstring theory and beyond}},
  Cambridge Monographs on Mathematical Physics. Cambridge University Press, 12,
  2007,
  \href{https://doi.org/10.1017/CBO9780511618123}{10.1017/CBO9780511618123}.

\bibitem{Cicoli:2008gp}
M.~Cicoli, C.~P. Burgess and F.~Quevedo, \emph{{Fibre Inflation: Observable
  Gravity Waves from IIB String Compactifications}},
  \href{https://doi.org/10.1088/1475-7516/2009/03/013}{\emph{JCAP} {\bfseries
  03} (2009) 013} [\href{https://arxiv.org/abs/0808.0691}{{\ttfamily
  0808.0691}}].

\bibitem{Klaewer:2016kiy}
D.~Klaewer and E.~Palti, \emph{{Super-Planckian Spatial Field Variations and
  Quantum Gravity}}, \href{https://doi.org/10.1007/JHEP01(2017)088}{\emph{JHEP}
  {\bfseries 01} (2017) 088}
  [\href{https://arxiv.org/abs/1610.00010}{{\ttfamily 1610.00010}}].

\bibitem{Bedroya:2019snp}
A.~Bedroya and C.~Vafa, \emph{{Trans-Planckian Censorship and the Swampland}},
  \href{https://doi.org/10.1007/JHEP09(2020)123}{\emph{JHEP} {\bfseries 09}
  (2020) 123} [\href{https://arxiv.org/abs/1909.11063}{{\ttfamily
  1909.11063}}].

\bibitem{Ooguri:2006in}
H.~Ooguri and C.~Vafa, \emph{{On the Geometry of the String Landscape and the
  Swampland}},
  \href{https://doi.org/10.1016/j.nuclphysb.2006.10.033}{\emph{Nucl. Phys. B}
  {\bfseries 766} (2007) 21}
  [\href{https://arxiv.org/abs/hep-th/0605264}{{\ttfamily hep-th/0605264}}].

\bibitem{Baume:2016psm}
F.~Baume and E.~Palti, \emph{{Backreacted Axion Field Ranges in String
  Theory}}, \href{https://doi.org/10.1007/JHEP08(2016)043}{\emph{JHEP}
  {\bfseries 08} (2016) 043}
  [\href{https://arxiv.org/abs/1602.06517}{{\ttfamily 1602.06517}}].

\bibitem{Ooguri:2018wrx}
H.~Ooguri, E.~Palti, G.~Shiu and C.~Vafa, \emph{{Distance and de Sitter
  Conjectures on the Swampland}},
  \href{https://doi.org/10.1016/j.physletb.2018.11.018}{\emph{Phys. Lett. B}
  {\bfseries 788} (2019) 180}
  [\href{https://arxiv.org/abs/1810.05506}{{\ttfamily 1810.05506}}].

\bibitem{Grimm:2018ohb}
T.~W. Grimm, E.~Palti and I.~Valenzuela, \emph{{Infinite Distances in Field
  Space and Massless Towers of States}},
  \href{https://doi.org/10.1007/JHEP08(2018)143}{\emph{JHEP} {\bfseries 08}
  (2018) 143} [\href{https://arxiv.org/abs/1802.08264}{{\ttfamily
  1802.08264}}].

\bibitem{Blumenhagen:2018nts}
R.~Blumenhagen, D.~Kl\"awer, L.~Schlechter and F.~Wolf, \emph{{The Refined
  Swampland Distance Conjecture in Calabi-Yau Moduli Spaces}},
  \href{https://doi.org/10.1007/JHEP06(2018)052}{\emph{JHEP} {\bfseries 06}
  (2018) 052} [\href{https://arxiv.org/abs/1803.04989}{{\ttfamily
  1803.04989}}].

\bibitem{Grimm:2018cpv}
T.~W. Grimm, C.~Li and E.~Palti, \emph{{Infinite Distance Networks in Field
  Space and Charge Orbits}},
  \href{https://doi.org/10.1007/JHEP03(2019)016}{\emph{JHEP} {\bfseries 03}
  (2019) 016} [\href{https://arxiv.org/abs/1811.02571}{{\ttfamily
  1811.02571}}].

\bibitem{Corvilain:2018lgw}
P.~Corvilain, T.~W. Grimm and I.~Valenzuela, \emph{{The Swampland Distance
  Conjecture for K\"ahler moduli}},
  \href{https://doi.org/10.1007/JHEP08(2019)075}{\emph{JHEP} {\bfseries 08}
  (2019) 075} [\href{https://arxiv.org/abs/1812.07548}{{\ttfamily
  1812.07548}}].

\bibitem{Lee:2018urn}
S.-J. Lee, W.~Lerche and T.~Weigand, \emph{{Tensionless Strings and the Weak
  Gravity Conjecture}},
  \href{https://doi.org/10.1007/JHEP10(2018)164}{\emph{JHEP} {\bfseries 10}
  (2018) 164} [\href{https://arxiv.org/abs/1808.05958}{{\ttfamily
  1808.05958}}].

\bibitem{Lee:2018spm}
S.-J. Lee, W.~Lerche and T.~Weigand, \emph{{A Stringy Test of the Scalar Weak
  Gravity Conjecture}},
  \href{https://doi.org/10.1016/j.nuclphysb.2018.11.001}{\emph{Nucl. Phys. B}
  {\bfseries 938} (2019) 321}
  [\href{https://arxiv.org/abs/1810.05169}{{\ttfamily 1810.05169}}].

\bibitem{Gendler:2020dfp}
N.~Gendler and I.~Valenzuela, \emph{{Merging the weak gravity and distance
  conjectures using BPS extremal black holes}},
  \href{https://doi.org/10.1007/JHEP01(2021)176}{\emph{JHEP} {\bfseries 01}
  (2021) 176} [\href{https://arxiv.org/abs/2004.10768}{{\ttfamily
  2004.10768}}].

\bibitem{Bastian:2020egp}
B.~Bastian, T.~W. Grimm and D.~van~de Heisteeg, \emph{{Weak gravity bounds in
  asymptotic string compactifications}},
  \href{https://doi.org/10.1007/JHEP06(2021)162}{\emph{JHEP} {\bfseries 06}
  (2021) 162} [\href{https://arxiv.org/abs/2011.08854}{{\ttfamily
  2011.08854}}].

\bibitem{Montero:2019ekk}
M.~Montero, T.~Van~Riet and G.~Venken, \emph{{Festina Lente: EFT Constraints
  from Charged Black Hole Evaporation in de Sitter}},
  \href{https://doi.org/10.1007/JHEP01(2020)039}{\emph{JHEP} {\bfseries 01}
  (2020) 039} [\href{https://arxiv.org/abs/1910.01648}{{\ttfamily
  1910.01648}}].

\bibitem{Montero:2021otb}
M.~Montero, C.~Vafa, T.~Van~Riet and G.~Venken, \emph{{The FL bound and its
  phenomenological implications}},
  \href{https://doi.org/10.1007/JHEP10(2021)009}{\emph{JHEP} {\bfseries 10}
  (2021) 009} [\href{https://arxiv.org/abs/2106.07650}{{\ttfamily
  2106.07650}}].

\bibitem{Calderon-Infante:2020dhm}
J.~Calder\'on-Infante, A.~M. Uranga and I.~Valenzuela, \emph{{The Convex Hull
  Swampland Distance Conjecture and Bounds on Non-geodesics}},
  \href{https://doi.org/10.1007/JHEP03(2021)299}{\emph{JHEP} {\bfseries 03}
  (2021) 299} [\href{https://arxiv.org/abs/2012.00034}{{\ttfamily
  2012.00034}}].

\bibitem{Grimm:2019ixq}
T.~W. Grimm, C.~Li and I.~Valenzuela, \emph{{Asymptotic Flux Compactifications
  and the Swampland}},
  \href{https://doi.org/10.1007/JHEP06(2020)009}{\emph{JHEP} {\bfseries 06}
  (2020) 009} [\href{https://arxiv.org/abs/1910.09549}{{\ttfamily
  1910.09549}}].

\bibitem{Nojiri:2002hz}
S.~Nojiri, S.~D. Odintsov and S.~Ogushi, \emph{{Friedmann-Robertson-Walker
  brane cosmological equations from the five-dimensional bulk (A)dS black
  hole}}, \href{https://doi.org/10.1142/S0217751X02012156}{\emph{Int. J. Mod.
  Phys. A} {\bfseries 17} (2002) 4809}
  [\href{https://arxiv.org/abs/hep-th/0205187}{{\ttfamily hep-th/0205187}}].

\bibitem{Chen:2014fqa}
S.~Chen, G.~W. Gibbons, Y.~Li and Y.~Yang, \emph{{Friedmann's Equations in All
  Dimensions and Chebyshev's Theorem}},
  \href{https://doi.org/10.1088/1475-7516/2014/12/035}{\emph{JCAP} {\bfseries
  12} (2014) 035} [\href{https://arxiv.org/abs/1409.3352}{{\ttfamily
  1409.3352}}].

\end{thebibliography}\endgroup

\end{document}